\documentclass[11pt,a4paper]{emulateapj}
\usepackage{epsfig}
\usepackage{amsmath}
\usepackage{natbib}

\begin{document}

\title{Two New Long--Period Hot Subdwarf Binaries with Dwarf Companions\footnotemark[*]}
\footnotetext[*]{Based on observations obtained with the Hobby--Eberly
  Telescope, which is a joint project of the University of Texas at
  Austin, the Pennsylvania State University, Stanford University,
  Ludwig-Maximilians-Universit\"{a}t M\"{u}nchen, and
  Georg-August-Universit\"{a}t G\"{o}ttingen.}
\author{Brad N. Barlow$^1$\footnotemark[\dag], Sandra E. Liss$^2$, Richard A. Wade$^1$, and Elizabeth M. Green$^3$}
\affiliation{\scriptsize $^1$Dept of Astronomy and Astrophysics, The Pennsylvania State University, 525 Davey Lab, University Park, PA 16802, USA\\ 
$^2$Department of Astronomy, University of Virginia, P.O. Box 400325, Charlottesville, VA 22904-4325, USA \\
$^3$Steward Observatory, University of Arizona, 933 N. Cherry Avenue, Tucson, AZ 85721, USA}\footnotetext[\dag]{\tt bbarlow@psu.edu}

\submitted{Accepted for publication in The Astrophysical Journal} 

\begin{abstract}
  Hot subdwarf stars with F--K main sequence binary companions have been
  known for decades, but the first orbital periods for such systems were
  published just recently.  Current observations suggest that most have long periods, on the order of years, and that some
  are or once were hierarchical triple systems.  As part of a survey
  with the Hobby--Eberly Telescope, we have been monitoring the radial
  velocities of several composite--spectra binaries since 2005 in order
  to determine their periods, velocities, and eccentricities.  Here we
  present observations and orbital solutions for
  two of these systems, PG 1449+653 and PG 1701+359.  Similar to the
  other sdB+F/G/K binaries with solved orbits, their periods are long, 909 d and 
  734 d, respectively, and pose a challenge to current binary
  population synthesis models of hot subdwarf stars.  Intrigued by their
  relatively large systemic velocities, we also present a kinematical
  analysis of both targets and find that neither is likely a member of
  the Galactic thin disk.
\end{abstract}

\section{Introduction}
\label{sec:intro}

Mystery has shrouded the formation of hot subdwarf stars since they were
first recognized as a blue extension of the horizontal branch in the
Galactic halo by \citet{gre74} almost forty years ago.  Since then, they
have been found in all Galactic
populations \citep{alt04,nap08}, globular clusters, and even the
elliptical galaxy M32 \citep{bro00}.  
 In this regard, it is interesting to note that the known hot
   subdwarfs seem to occur only in rather old populations, with ages
   greater than about 4--5 Gyr, and are much more common (as a percentage
   of the core He-burning stars) in populations with high metallicity,
   e.g., the old, metal-rich open clusters NGC 188 and NGC 6791 \citep{lie94, gre06}.

Models show hot subdwarf B (sdB)
stars are core He-burning stars surrounded by extremely thin H envelopes,
 with masses around 0.5 M$_{\sun}$ and
radii near 0.2 R$_{\sun}$ (\citealt{heb86,saf94}).  To explain their
lack of hydrogen (in contrast to other core helium--burning stars), 
various single--star and binary--evolution
formation scenarios have been constructed, all of which entail the sdB
progenitor losing its outer hydrogen envelope while on or near the red
giant branch (RGB).  As single--star scenarios are {\em ad hoc},
binary--star hypotheses have generally been favored to explain the loss 
of the H envelope, following early theoretical
work by \citet{men76}.

\citet{han02,han03} presented the first results from a major binary
population synthesis (BPS) study of hot subdwarf stars and described
five primary formation channels, including two common envelope (CE)
channels, two stable Roche lobe overflow (RLOF) channels, and the merger
of a He--core white dwarf binary.  Other notable studies include those
of \citet{cla11}, who demonstrate that the merger of a He--core white
dwarf and M dwarf can create an sdB, and \cite{nel10}, who put forth an
alternative CE channel employing the $\gamma$--formalism.  Many
assumptions influence the outcomes
in BPS models, including the common--envelope ejection efficiency,
minimum mass for core He ignition, and envelope binding energy.
  Changing any one of these parameters,
even moderately, can greatly alter the
hot subdwarf population predicted by the codes.  \citet{cla12} recently
demonstrated this sensitivity using a grid of BPS models with different
inputs.  They found that (1) a wide range of parameter sets can
reproduce the observed subpopulation of short--period white dwarf and M
dwarf binaries, which are products of CE evolution, and (2) observations
of systems with F--K main sequence (MS) companions are needed to constrain
the physics governing RLOF processes.

While there are numerous short--period orbits (days) published for
post--CE sdB binaries (see Table A.1 of \citealt{muchfuss}), no definitive periods
under 30 d have been reported for systems with MS companions earlier
than spectral type M.\footnote{\citet{mon10}
  suggest that the star M5865 in globular cluster NGC 6752 is a hot
  subdwarf + G--K MS binary with a period of order 5 d, but details have
  not been published.} On the
contrary, long--term monitoring of these systems
\citep{dec12,ost12,bar12,vos12} has shown their periods are on the order of
years, as originally suspected by \citet{saf01} more than a decade ago. Such long periods appear
 to be inconsistent with the predictions of \citet{han02,han03} and pose a major challenge
to BPS models.  Of course, orbital parameters for many more long--period
systems are needed before any definitive conclusions can be drawn.

In 2005 we began monitoring the radial velocities of 15 hot subdwarf
stars with F--K main sequence companions at the Hobby--Eberly Telescope
(HET) in order to constrain their orbital parameters.  Details of this
study may be found in \citealt{bar12} (hereafter, Paper I).  Although we
could only solve for the orbits of three systems using our initial data
set, we presented preliminary periods for 12 targets.  Almost all have
periods in excess of 100 days; only two appear to have short periods,
and these targets are likely hierarchical triple-- or quadruple--star systems.  Our
preliminary results confirm previous findings that composite--spectra
sdB systems tend to have long periods.  To solve for the
orbits of the remaining systems, we began collecting follow--up
observations with the High Resolution Spectrograph on the HET in Mar
2012.  An up--to--date orbital period histogram of all hot subdwarf
binaries (Fig. 5 of Paper I) shows a clear dichotomy between systems
with F-K main sequence companions (P$>$ 100 d) and those with white dwarf
or M dwarf companions (P$<$ 100 d).  A possible period gap is emerging
around P$\sim$100 d, but the statistics are presently too
poor to claim this feature as real.

Here we present new spectroscopic observations of the sdB+MS binaries 
PG 1449+653 (V=13.6) and PG 1701+359 (V=13.2).  We find orbital periods
exceeding two years for both targets, a result
in line with other sdB+F/G/K binaries.  Intrigued by their relatively
large systemic velocities, we combine our spectroscopy with proper
motion measurements to calculate Galactic orbits and find that neither
system is likely a member of the Galactic thin disk.

\section{Observations \& Reductions}
\label{sec:results}

From 2005 to 2008 we monitored PG 1449+653 and PG 1701+359 
spectroscopically with the Medium--Resolution Spectrograph (MRS; 
\citealt{ram98}) on the Hobby--Eberly Telescope (HET).  Full details on 
the observational setup used may be found in Paper I.  We commenced 
follow--up observations in 2012 with the High Resolution Spectrograph 
(HRS) on the HET, using the 2\arcsec optical fiber pair, the 316 mm$^{-1}$ 
cross-disperser grating, and 2$\times$3 on--chip binning to achieve an 
average resolution of R=30,000 over the 4076--7838 \AA\ spectral range. 
This configuration yields an average dispersion of $\sim$ 3.6 km s$^{-1}$ 
per binned pixel and samples $\sim$ 2.7 pixels per resolution element.  
Target spectra were taken in pairs and later combined to help remove 
cosmic rays. Standard calibration frames were also collected each night 
using the same instrumental setup, including twilight sky spectra, bias 
frames, quartz lamp flatÐfield spectra, and, on most occasions, spectra 
of telluric, RV, and spectrophotometric standard stars. 

We bias--subtracted and flat--fielded all HET data using the \textit{ccdproc}
routine in {\sc IRAF}\footnote{IRAF is distributed by the National Optical 
Astronomy Ob- servatories, which are operated by the Association of 
Universities for Research in Astronomy, Inc., under cooperative agreement 
with the National Science Foundation} and optimally extracted apertures 
with the \textit{apall} function.  Some echelle orders were discarded due 
to CCD fringing, heavy contamination from sky emission and telluric 
absorption features, cross--disperser order overlap, and low throughput.  
For HRS (MRS), we kept a total of 48 (14) orders covering 4100--7000 
\AA\ (4400--6200 \AA).  

We supplement the HET data with single--order, long--slit spectra from the Blue Spectrograph at
the Multi Mirror Telescope (MMT) obtained in 1997 and from 2002 to
2003.  An 832 mm$^{-1}$ grating was used in second
order to achieve wavelength coverage of 4000--4950 \AA\ at a
resolution of 1.1 \AA\ (R = 4100) and dispersion near 0.36 \AA\
pixel$^{-1}$.  MMT data were reduced using the same IRAF routines
mentioned above for the HET observations.  Table \ref{tab:log}
summarizes the entire data set.  
 \begin{table}
 \centering
  \caption{Summary of Spectroscopic Observations} 
  \begin{center}
    \begin{tabular}{llllll} \hline \hline              
Target & Telescope/Instrum. & N$_{\rm obs}$ & Year\\
\hline
PG  1449+653 &HET/HRS & 7 & 2012--2013\\
& HET/MRS & 8 & 2005--2008\\

 & MMT/Blue Spect. & 5 & 1997\\
 
PG 1701+359 & HET/HRS &  8   & 2012--2013\\
			& HET/MRS & 13 & 2005--2008\\
 & MMT/Blue Spect. & 8 & 1997, 2002--2003\\
\hline
    \end{tabular}
  \end{center}
  \label{tab:log}
\end{table}

\section{Analysis}
\label{sec:results}
We generally followed the same procedures outlined in Paper I to measure
radial velocities (RVs) and determine the best-fitting orbital
parameters for each system; we highlight only certain aspects of these
procedures here (along with any deviations from Paper I) and refer
readers to Paper I for full disclosure of our analysis methods.

 \begin{table*}
 \centering
  \caption{Heliocentric Radial Velocities of the Cool Companion} 
  \begin{center}
    \leavevmode
    \begin{tabular}{c c l c c c l} \hline \hline           
    HJD & RV$_{\rm MS}$ & Facility & & HJD & RV$_{\rm MS}$ & Facility\\
    -2450000 & [km s$^{-1}$] &  && -2450000 & [km s$^{-1}$] & \\
    \hline   
    \multicolumn{3}{c}{PG 1449+653} && \multicolumn{3}{c}{PG 1701+359}\\
       \multicolumn{3}{c}{------------------------------------------------} && \multicolumn{3}{c}{------------------------------------------------}\\
    511.0187 &   -128.4 $\pm$   1.0 & MMT & &      627.8929  &  -121.0 $\pm$   1.9 & MMT \\ 
      626.8101 &   -129.3 $\pm$   1.3 & MMT & &      633.8716  &  -123.3 $\pm$   1.9 & MMT \\ 
      627.7691 &   -129.1 $\pm$   1.0 & MMT & &      642.8422  &  -122.2 $\pm$   1.9 & MMT \\ 
      642.7782 &   -129.3 $\pm$   1.0 & MMT & &      643.7559  &  -124.8 $\pm$   1.8 & MMT \\ 
      701.6427&   -130.6 $\pm$   1.4 & MMT & &      701.6934  &  -122.9 $\pm$   2.0 & MMT \\ 
     3462.8508 &   -136.0 $\pm$   0.8 & HET/MRS & &     1654.7222  &  -119.6 $\pm$   1.1 & MMT \\ 
     3479.8705 &   -136.4 $\pm$   1.0 & HET/MRS & &     2544.6458  &  -123.0 $\pm$   1.7 & MMT \\ 
     3503.8035 &   -138.0 $\pm$   0.9 & HET/MRS & &     2826.8606  &  -123.9 $\pm$   1.6 & MMT\\ 
     3520.6751 &   -138.4 $\pm$   0.8 & HET/MRS & &     3482.8107  &  -122.2 $\pm$   1.2 & HET/MRS \\ 
     3758.0254 &   -143.5 $\pm$   0.7 & HET/MRS & &     3520.9447  &  -122.4 $\pm$   0.9 & HET/MRS \\ 
     3827.9095 &   -138.4 $\pm$   1.1 & HET/MRS & &     3543.8831  &  -123.8 $\pm$   1.3 & HET/MRS \\ 
     3855.8400 &   -138.3 $\pm$   1.2 & HET/MRS & &     3584.7790  &  -123.6 $\pm$   1.1 & HET/MRS \\ 
     4216.8374 &   -128.5 $\pm$   0.8 & HET/MRS & &     3609.7073  &  -122.8 $\pm$   1.6 & HET/MRS \\ 
     5991.9222 &   -128.4 $\pm$   0.2 & HET/HRS & &     3827.8767  &  -120.4 $\pm$   1.3 & HET/MRS \\ 
     6018.8438 &   -128.5 $\pm$   0.2 & HET/HRS & &     3854.7918  &  -119.0 $\pm$   1.3 & HET/MRS \\ 
     6061.6989 &   -129.5 $\pm$   0.2 & HET/HRS & &     4273.8819  &  -122.1 $\pm$   1.2 & HET/MRS \\ 
     6086.6883 &   -130.2 $\pm$   0.2 & HET/HRS & &     4506.0149  &  -121.3 $\pm$   0.9 & HET/MRS \\ 
     6118.6349 &   -131.1 $\pm$   0.2 & HET/HRS & &     4569.8307  &  -117.1 $\pm$   2.9 & HET/MRS \\ 
     6334.9466 &   -142.6 $\pm$   0.2 & HET/HRS & &     4582.7928  &  -118.8 $\pm$   0.8 & HET/MRS \\ 
     6376.9266 &   -144.2 $\pm$   0.2 & HET/HRS & &     4612.7189  &  -117.1 $\pm$   1.5 & HET/MRS \\ 
       ... &    ...  & ... & &     4671.7919  &  -115.3 $\pm$   1.4 & HET/MRS \\ 
       ... &    ... & ...  & &     5997.8963  &  -121.0 $\pm$   0.4 & HET/HRS \\ 
       ... &    ...  &...  & &     6040.7881  &  -119.9 $\pm$   0.2 & HET/HRS \\ 
       ... &    ... & ...  & &     6068.9367  &  -118.7 $\pm$   0.3 & HET/HRS \\ 
       ... &    ...  & ...  & &     6098.6339  &  -117.6 $\pm$   0.2 & HET/HRS \\ 
       ... &    ... & ...  & &     6145.7336  &  -116.9 $\pm$   0.2 & HET/HRS \\ 
       ... &    ... & ...  & &     6191.6229  &  -116.2 $\pm$   0.3 & HET/HRS \\ 
       ... &    ... & ...  & &     6326.0072  &  -118.3 $\pm$   0.3 & HET/HRS \\ 
       ... &    ...  &  ... & &     6370.8962  &  -119.5 $\pm$   0.3 & HET/HRS \\ 
\hline
    \end{tabular}
  \end{center}
  \label{tab:RVs}
\end{table*}

\subsection{Companion classification}
\label{subsec:classification}
We cross--correlated the PG 1449+653 and PG 1701+359 spectra with MS
standards (F0--K7) observed by the MMT and HET in the same instrumental
configurations as our data.  {\sc iraf}'s \textit{fxcor} routine was
used for all regions of the spectra except those around the hydrogen
Balmer and helium lines.  We determined the best--matching MS spectra
for each target from the Tonry--Davis ratio (R--value output from \textit{fxcor}; \citealt{ton79}) and the
\textit{fxcor}--outputted velocity errors for each template.  A spectral
type of G0V was consistently found for PG 1449+653.  The PG 1701+359
spectrum, which has lower S/N and higher dilution, was best--matched to
a K0V standard, although several G--type standards (as early as G0V)
also fit the spectra reasonably well.

We also adopt the approach used by \citet{sta03} to photometrically
determine the spectral types of the cool companions.  After correcting
for reddening \citep{sch98}, we compare the colors of PG 1449+653 and PG
1701+359 to a theoretical grid of $B$-$V$ and $J$-$K$ colors for hot
subdwarf binaries with cool main sequence companions.  We note that 
Population I colors were assumed for the cool companions,
although kinematics suggest these systems might have halo--like orbits
(\S \ref{sec:orbits}).  For each pairing, the dilution factor of the
companion spectrum in the $V$--band ($D_V \equiv L_{\rm comp} / L_{\rm
  total}$) was varied from 1--100\%.  Color indices for the cool
companions and hot subdwarfs are taken from \citet{joh66} and
\citet{sta03}, respectively.  For PG 1449+653, we find colors consistent
with a 'typical' hot subdwarf star and a F8--G3 dwarf (best fit:
G1) with a dilution factor around $D_V$ $\sim$ 0.3; these results
agree with the spectroscopically--determined value.  PG 1701+359's
composite colors imply a G7--K2 dwarf companion (best fit: K0)
with $D_V \sim 0.15$, also consistent with the observed composite
spectrum.

\subsection{Radial velocity measurements}
Heliocentric radial velocities were determined by cross--correlating
each spectrum against velocity standards using {\sc IRAF}'s
\textit{fxcor} and \textit{rvcorrect} routines.  All MMT spectra were
correlated against the velocity standard HD 39587, a G0V dwarf, which
was observed using the same instrumental setup as our target
observations.  For the HET data, we achieve the best results when using
a single, high S/N twilight sky spectrum as the correlation template for
all HRS and MRS observations.  We extracted HET velocities from each
order individually, being sure to avoid spectral regions where subdwarf
features compromise our ability to measure the cool companion
velocities.  We adopt as the error the standard deviation of the mean of the
order--by--order velocity fits.  Typical velocity errors
ranged from 1--1.5 km~s$^{-1}$ for MRS, 200-400 m~s$^{-1}$ for HRS, 
and 1--1.5 km s$^{-1}$ for the MMT spectra. Table
\ref{tab:RVs} presents all cool companion velocities and their
associated errors. 

 \begin{table}
 \centering
  \caption{Heliocentric Radial Velocities\\ of the sdB in PG 1449+653} 
  \begin{center}
    \leavevmode
    \begin{tabular}{c c c } \hline \hline           
    HJD & RV$_{\rm MS}$ & Facility \\
    -2450000 & [km s$^{-1}$] & \\
    \hline   
       3462.8508 &       -129.9 $\pm$       5.0 & HET/MRS\\
       3479.8705 &       -130.5 $\pm$       6.0 & HET/MRS\\
       3503.8035 &       -130.3 $\pm$       5.6  & HET/MRS\\
       3520.6751 &       -124.5 $\pm$       5.0  & HET/MRS\\
       3758.0254 &       -122.3 $\pm$       3.9 & HET/MRS\\
       3827.9095 &       -122.5 $\pm$       6.6  & HET/MRS\\
       3855.8400 &       -128.3 $\pm$       7.2  & HET/MRS\\
       4216.8374 &       -135.3 $\pm$       4.7  & HET/MRS\\
       5991.9222 &       -146.9 $\pm$       3.3  & HET/HRS\\
       6018.8438 &       -142.5 $\pm$       2.9 & HET/HRS\\
       6061.6989 &       -144.4 $\pm$       3.8  & HET/HRS\\
       6086.6883 &       -140.8  $\pm$       2.9 & HET/HRS\\
       6118.6349 &       -137.7  $\pm$       2.9 & HET/HRS\\
       6334.9466 &       -120.2  $\pm$       3.0 & HET/HRS\\
       6376.9266 &       -119.9 $\pm$       3.1 & HET/HRS\\

\hline
    \end{tabular}
  \end{center}
  \label{tab:RVs_sdB_1449}
\end{table}

  \begin{figure*}
  \begin{center}
   \includegraphics{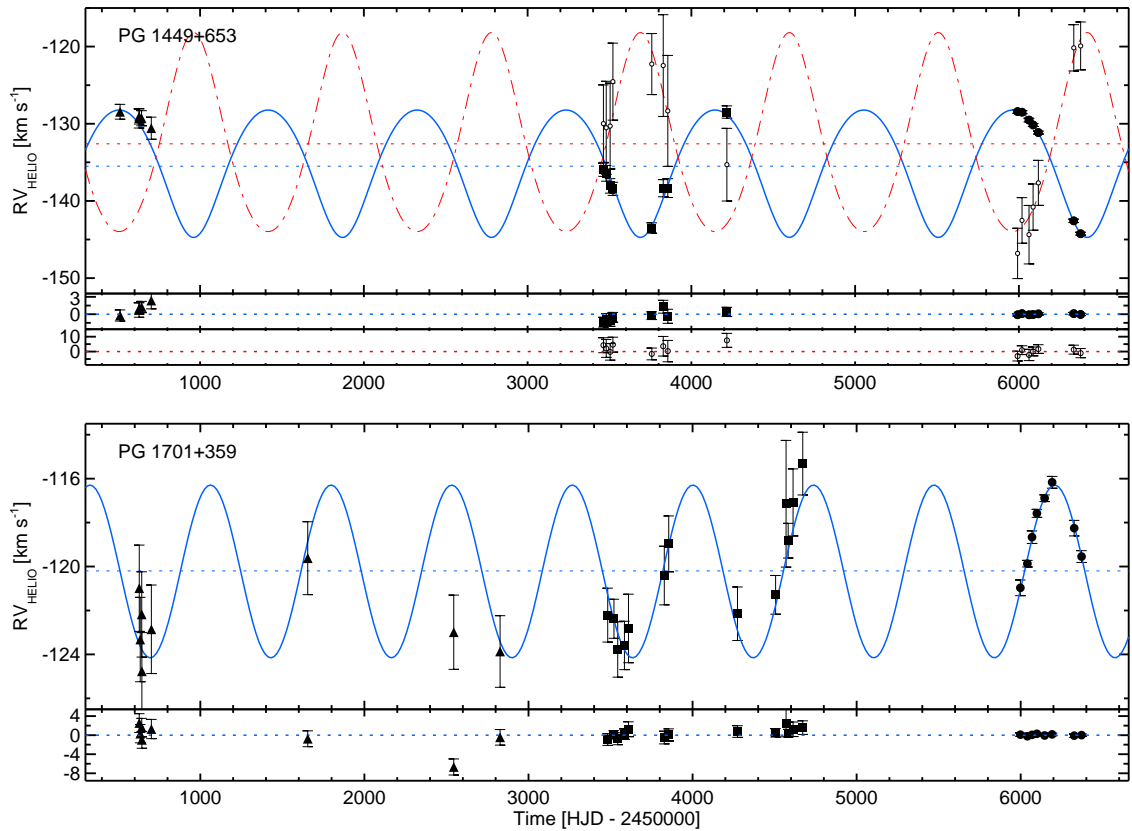}
   \caption{Heliocentric radial velocities of PG 1449+653 (top panel) 
   and PG 1701+359 (bottom panel). Cool companion measurements 
   from HET/MRS, HET/HRS, and MMT/Blue Spectrograph are shown 
   with filled circles, squares, and triangles, respectively. Open circles in the top panel mark 
   the sdB velocities measured from He lines in the HET spectra. 
   Some of the error bars are difficult to see as they are smaller than the symbols. The
   best--fitting circular orbital solutions for the sdB and cool companion are 
   denoted with dot--dashed (orange) and solid (blue) lines, respectively. 
   Residuals from this fit are shown in the lower portions of the panels 
   for the cool companion and, for PG 1449+653, the hot subdwarf. (A color version of this figure is 
   available in the online journal.) }
     \label{fig:rvs}
  \end{center}
\end{figure*}

We attempted to measure the orbital reflex motion of the hot subdwarf in
each system by examining the positions of \ion{He}{1} \& \ion{He}{2}
absorption profiles.  We avoided using the sdB's \ion{H}{0}Balmer lines
as they are contaminated by narrower Balmer lines from the cool
companion (and in the MRS/HRS spectra, they are
broader than the wavelength coverage of individual echelle orders).
For PG 1449+653, we extracted RVs by cross--correlating the \ion{He}{1} 5875
\AA\ profile in each spectrum against a self template, chosen to be the
highest S/N spectrum obtained.  The template's zero point was determined
by cross--correlating the template against an sdB model spectrum with
T$_{\rm eff}$ = 30000 K and $\log g = 5.58$.  This synthetic template
includes pressure broadening and asymmetry in the 5875 \AA\ \ion{He}{1} 
multiplet.  Table \ref{tab:RVs_sdB_1449} summarizes the sdB velocities
measured from all HET spectra.  Unfortunately, our efforts to extract
precise sdB velocities were futile in the case of PG 1701+359.  
This shortcoming likely results from a combination of a lower S/N in many of the spectra 
(due to shorter exposure times) and intrinsically weaker \ion{He}{1} lines, 
which could result from the sdB in PG 1701+359 having a significantly different photospheric helium abundance 
and/or effective temperature than the sdB in PG 1449+653.  We also note that the 
orbital velocities associated with 
PG 1701+359 are more than a factor of two smaller than those of PG 1449+653 and,
naturally, harder to detect.


 \begin{table*}
 \centering
  \caption{Orbital Parameters} 
  \begin{center}
    \leavevmode
    \begin{tabular}{c rl rl rl rl rl rl rl} \hline \hline              
Target & \multicolumn{2}{c}{$P$} &  \multicolumn{2}{c}{$T_{0}$$^a$} &  \multicolumn{2}{c}{$e$} &  \multicolumn{2}{c}{$\omega$} &  \multicolumn{2}{c}{$K_{\rm MS}$} &    \multicolumn{2}{c}{$K_{\rm sdB}$} &\multicolumn{2}{c}{$\gamma$}\\
&  \multicolumn{2}{c}{[d]}  &  \multicolumn{2}{c}{[HJD-2450000]} & \multicolumn{2}{c}{}  &  \multicolumn{2}{c}{[deg]} &  \multicolumn{2}{c}{[km s$^{-1}$]} &  \multicolumn{2}{c}{[km s$^{-1}$]} & \multicolumn{2}{c}{[km s$^{-1}$]}\\
\hline
PG 1449+653 & 909 &  $\pm$ 2 & 3675 & $\pm$ 35 & 0.11 & $\pm$ 0.02 & 174 & $\pm$ 15& 8.2 & $\pm$ 0.3 & 12.8 & $\pm$ 1.1  & -135.5 & $\pm$ 0.2 \\  
                           &  908 & $\pm$ 2 & 3251 & $\pm$ 40&  \multicolumn{2}{c}{0.0 (fixed)} &  \multicolumn{2}{c}{...}& 8.3 & $\pm$ 0.2 & 13.5 & $\pm$ 0.8 & -136.4 & $\pm$ 0.2\\  

PG 1701+359 & 738 & $\pm$ 4 & 3869 & $\pm$ 80 & 0.07 & $\pm$ 0.04 & 298 & $\pm$ 38 & 3.6 & $\pm$ 0.2 &  \multicolumn{2}{c}{...} & -120.1 & $\pm$ 0.2\\   
                          & 734 & $\pm$ 3&  3269 & $\pm$ 60&  \multicolumn{2}{c}{0.0 (fixed)} &  \multicolumn{2}{c}{...} & 3.9& $\pm$ 0.2 &  \multicolumn{2}{c}{...}  & -120.2 & $\pm$ 0.2\\  
\hline
\multicolumn{15}{l}{\scriptsize $^a$Time of velocity maximum (maximum redshift) in the cool companion RV curve ($e = 0$ case) or time of periastron passage ($e \ne 0$).}\\
    \end{tabular}
  \end{center}
  \label{tab:orbital_parameters}
\end{table*}



\subsection{Determining the orbital parameters}
We solved for the full set of orbital parameters ($P$, $K$, $e$,
$\omega$, $\gamma$, $T_{\rm 0}$) using the {\sc IDL}--based {\sc rvlin}
software \citep{wri09} and estimated parameter uncertainties using a
{\sc rvlin}--drive bootstrapping technique \citep{wan12}.  
After first running the code with eccentricity ($e$) fixed to
zero, we refitted the data with it and the argument of periapsis
($\omega$) left adjustable.  We also allowed {\sc rvlin} to freely fit
any zero--point offsets between the MMT, MRS/HET, and HRS/HET
velocities, but as no significant shifts were found, we ultimately set
the offsets to zero.  To be certain {\sc rvlin} converged on the global
minimum, we computed a `floating--mean periodogram' \citep{cum99} for
each RV curve to investigate $\chi^2$ as a function of period.  During
this process, a series of sine waves are fitted to the data with
adjustable amplitudes and periods ranging from 0.1 to 10000 d (in steps
of log $P$[d] = 0.001).  Only one probable period emerges from the
periodogram for each target; the next--largest alias peaks are 10$^3$
times less probable.  These periods also correspond to the global minima
found by {\sc rvlin}, and so we are confident we have found the correct
periods. 

\vspace{-2mm}
\section{The Orbital Solutions}
 Figure \ref{fig:rvs} and Table \ref{tab:orbital_parameters} present 
the velocity curves and best--fitting orbital parameters for PG 1701+359 
and PG 1449+643, which we discuss in detail below.

 \begin{table*}
 \centering
  \caption{Coordinates, Velocities, and Distance Estimates} 
\scriptsize
  \begin{center}
    \leavevmode
    \begin{tabular}{ccccccc} \hline \hline              
Target & RA & Dec & $\mu_{\alpha} \cos \delta^a$ & $\mu_{\delta}^a$ & $\gamma$ & d$^{b}$\\
 & [J2000] & [J2000] & [mas yr$^{-1}$] & [mas yr$^{-1}$] & [km s$^{-1}$]  & [kpc] \\ 
\hline
PG 1449+653 &14:50:36.1 &+65:05:53&-21.7 $\pm$ 2.1 & 13.6 $\pm$ 1.0  &-135.5 $\pm$ 0.2 &0.98 $\pm$ 0.16\\     
PG 1701+359 &17:03:21.6 &+35:48:49&-57.9 $\pm$ 4.0   & 	20.4 $\pm$ 0.9& -120.2 $\pm$ 0.2& 0.69 $\pm$ 0.15\\  
\hline
PG 1104+243 & 11:07:26.3 & +24:03:12 &   -65.9 $\pm$ 1.2 &-25.1 $\pm$ 1.2  &      -15.68 $\pm$ 0.05& 0.33 $\pm$ 0.11\\  
PG 1317+123 & 13:19:53.6 & +12:03:59 &    -6.9  $\pm$ 1.1 &-1.6  $\pm$ 1.1 &    +40.3 $\pm$ 0.2& 0.33 $\pm$ 0.11\\   
PG 1338+611 & 13:40:14.7 & +60:52:48 &    14.5 $\pm$ 0.9  & -61.4 $\pm$ 0.8 &   +32.58 $\pm$ 0.07 & 0.32 $\pm$ 0.11\\  
\hline
\multicolumn{7}{l}{$^a$proper motions taken from UCAC4 \citep{zac12}}\\
\multicolumn{7}{l}{$^b$calculated from the distance modulus assuming a 'typical' sdB star and our best estimate of the}\\
\multicolumn{7}{l}{\hspace{1mm}companion's spectral type; error bars are conservative.}
    \end{tabular}
  \end{center}
  \label{tab:parameters}
\end{table*}

\subsection{PG 1449+653}
The abundance of absorption features from the early G dwarf 
provided robust RV measurements with mean 
errors ranging from 245 m s$^{-1}$ (HRS/HET) to 1400 m s$^{-1}$ (MMT).  
Measuring the reflex motion of the sdB proved more difficult as most of
the subdwarf's absorption features (H Balmer lines, \ion{He}{1} lines) were heavily contaminated
by features from the cool companion and could not provide reliable velocity estimates. 
 We were only able to extract velocities from the HET data, using
the 5875 \AA\ \ion{He}{1} line alone.  This feature, which is significantly diluted
due to continuum from the companion, appears in two adjacent 
orders in the MRS spectra, but only in one order in the HRS data.    
Using the methods described in \S 3.2, we were able to extract 
sdB RVs with errors ranging from 3--7 km s$^{-1}$.

The top panel of Figure \ref{fig:rvs} presents the MS and sdB velocity 
curves, which are 180 degrees out of phase, as expected for a binary 
system.  Given the relatively large uncertainties on the sdB measurements, 
we used only the MS velocities to determine the orbital 
parameters with {\sc rvlin}.    Table \ref{tab:orbital_parameters} lists 
the results for the best--fitting circular and elliptical orbits.  To determine whether the eccentric solution
is preferred, we apply the 
revised Lucy--Sweeney (LS) test \citep{luc12,luc71}, which uses Bayes' theorem
to determine bounds on the exact eccentricity using the value $e \pm \mu$ 
we measured from RVLIN.
According to the LS test, we can reject the circular orbit solution in favor of an eccentric
orbit with $e = 0.11 \pm 0.03$.   
For the remainder of the text, we continue under the assumption of an eccentric orbit, for
which we report $P$ = 909 $\pm$ 2 d. 


To derive the mass ratio, we also measured the semi--amplitude of the sdB RV
curve.  We fixed the period, eccentricity, and phase to the values determined from the cool companion
measurements (the phase was shifted by 180 deg) and allowed the
semi--amplitude and systemic velocity to float freely.  
Allowing $\gamma$ to adjust accounts for a difference in the surface gravities (gravitational redshifts) of the two
components, which results in apparent systemic velocities that differ from one another.
In sdB+F/G/KV systems, this difference can
reach upwards of 2 km s$^{-1}$ and was first measured for a hot subdwarf binary
by \citet{vos12}.  We find $K_{\rm sdB}$ = 12.8 $\pm$ 1.0 km
s$^{-1}$ and $\gamma_{\rm sdB}$ = -132.6 $\pm$ 1.1 km s$^{-1}$ for the sdB star.
The measured difference in systemic velocities is $\Delta \gamma$ = 2.9 $\pm$ 1.1 km s$^{-1}$.
This measurement, when combined with a log \textit{g} estimate for the main sequence companion, 
allows one to compute the gravitational redshift (and thus, log \textit{g}) 
for the hot subdwarf star, but in this case the errors are too large to make this a fruitful exercise.

From K$_{\rm MS}$ and K$_{\rm sdB}$, we calculate a mass
ratio of $q$ = $0.64$ $\pm$ $0.06$, where $q \equiv K_{\rm MS}/K_{\rm sdB}$. 
 If we assume $M_{\rm MS}$ = $1.04$
$\pm$ $0.12$ M$_{\sun}$ for the cool companion (based on our spectroscopic 
classification, which assumes solar abundances), we derive a subdwarf mass of $M_{\rm sdB}$ = 0.66 $\pm$
0.10 M$_{\sun}$.  With this precision, we cannot claim a significant
deviation from the canonical mass value ($\sim$0.5 M$_{\sun}$). 

\subsection{PG 1701+359}
Heavy dilution of the cool companion's spectral features from the hot
subdwarf's continuum made classifying the PG 1701+359 spectrum and
measuring velocities more difficult than in the case of PG 1449+653.
Errors on the cool companion RVs ranged from 250 m s$^{-1}$ (HRS) to
1--2 km s$^{-1}$ (MRS, MMT).  We were not able to derive useful sdB
velocities from the \ion{He}{1} lines.  In view of the expected
acceleration, which is much smaller than in the PG 1449+653 case, they
are too weak for our cross--correlation measurement technique to achieve
the required precision.


The cool companion RV curve (bottom panel, Figure \ref{fig:rvs})
exhibits a small--amplitude variation ($\sim$4 km s$^{-1}$) with a
period near 2.0 years.   Although an eccentric solution models our
data more accurately (naturally), we cannot yet rule out a circular orbit 
since the best-fitting $e = 0.07 \pm 0.04$ solution was rejected by the revised LS
test.  Instead, we replace our measured value with an upper limit, $e \le 0.23$, 
which was determined using their recommended prescription (Appendix 2 of \citealt{luc12}).
The derived orbital parameters from the
circular and non--circular models agree with one another to within the
errors.  
 Our current phase coverage is incomplete as nearly all of our
velocities are on the ascending portion of the RV curve; additional
measurements on the descending slope will greatly improve the
eccentricity estimate.  We continue under the assumption of $e = 0$, for which we report $P = 734 \pm 3$.

 \begin{table*}
 \centering
  \caption{Galactocentric 6-D Phase Space Coordinates and Orbital Parameters} 
\scriptsize
  \begin{center}
    \leavevmode
    \begin{tabular}{c|ccc|ccc|ccc|cccc} \hline \hline              
Target & $X$ & $Y$ & $Z$ & $U$ & $V$ & $W$ & $\Phi$ & $\Theta$ & $I_z$ &$R_a$ & $R_p$ & $z_{\rm max}$ & $e$\\
             & [kpc] & [kpc] & [kpc] & [km s$^{-1}$] & [km s$^{-1}$]& [km s$^{-1}$] &  [km s$^{-1}$]   & [km s$^{-1}$]& [kpc km s$^{-1}$]  & [kpc] & [kpc] &[kpc] &  \\
\hline
PG 1449+653 &   -8.67    & $+$0.64   &  $+$0.73  & -78   &$+$101  & -81 & $+$10  & $+$95 &   -829 &   2.61&     9.91     &3.73 &    0.58  \\  
PG 1701+359 &   -8.21   &  $+$0.46&     $+$0.41  &-132 &   $+$68  &  $+$94 &  $+$11   & $+$60 & -492    &1.48   & 10.69  &   4.70  &   0.76  \\     
\hline
PG 1104+243  &   -8.61 &  -0.07     &$+$0.30&   -59   &$+$168 &  -51  & $+$9.9  &$+$168  &   -1447  &   5.25 &    9.37    & 1.12  &    0.28 \\  
PG 1317+123  &  -8.42  &  -0.05     &$+$0.31    &$+$11   &$+$219   & $+$46  & $+$8.7   & $+$219 & -1843&     8.23&     8.88 &     0.87  &   0.04 \\  
PG 1338+611 &  -8.56    & $+$0.14     &$+$0.22    &$+$62   &$+$216    &$+$72 & $+$7.9   &$+$210&  -1797 &  6.96    &11.16 &    2.21     &0.23  \\   
\hline
    \end{tabular}
  \end{center}
  \label{tab:6D_coords}
\end{table*}

  \begin{figure*}
  \begin{center}
   \includegraphics[scale=0.85]{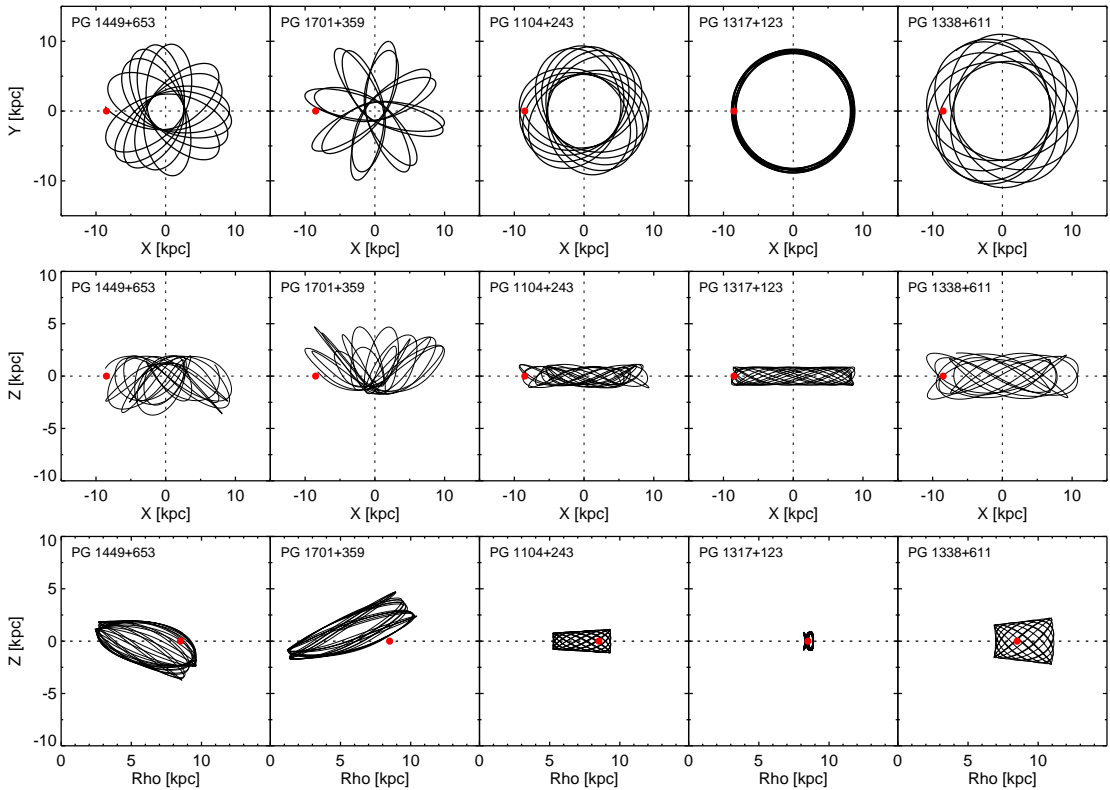}
   \caption{Galactocentric orbits for PG 1449+653, PG 
   1701+359, and the three binaries solved in 
   Paper I, shown over a 10 Gyr period.  Orbits were calculated using the 
   potential of \citet{all91}.  The top two rows show the orbits in the $Y$-$X$ 
   and $Z$-$X$ planes, while the bottom row shows the height above 
   the disk ($Z$) as a function of Galactocentric distance (Rho).  In each 
   panel, the location of the Sun is marked with a solid red circle, for reference.
    (A color version of this figure is 
   available in the online journal.)}
     \label{fig:orbits}
  \end{center}
\end{figure*}

\section{Galactic Kinematics}
\label{sec:orbits}


Both PG 1449+653 and PG 1701+359 display relatively large systemic
velocities, so we have calculated their Galactic trajectories to see
what conclusions can be drawn concerning their kinematic population
(halo vs.\ thick disk vs.\ thin disk).
We also include as a
comparison an analysis of the three systems solved in Paper I (PG
1104+243, PG 1317+123, \& PG 1338+611), which were found to have less
interesting space motions.  A kinematical study requires the full set of
6--D phase space coordinates, which we determine from the right
ascension ($\alpha$), declination ($\delta$), distance ($d$), systemic
velocity ($\gamma$), and proper motion ($\mu_{\alpha} \cos \delta$,
$\mu_{\delta}$).  We summarize these input parameters in Table
\ref{tab:parameters}.  

We adopt the values of $\gamma$ shown in Table
\ref{tab:orbital_parameters} as the systemic
velocity inputs ($e$=0.11 solution for PG 1449+653, $e$=0 solution for PG 1701+359).  Proper motion measurements with errors ranging from
 3--11 mas yr$^{-1}$ were taken from the Fourth U.S.
Naval Observatory CCD Astrograph Catalog (UCAC4; \citealt{zac12}).  The
most difficult parameter to determine is the distance.  Here we estimate
a range of probable line--of--sight distances to each binary from the
distance moduli.  We assume $T_{\rm eff}$ = 24000--38000 K and $R$ =
0.15--0.22 R$_{\sun}$ for the sdB, taking into account conventional
correlations between these parameters for hot subdwarfs.  For the cool
companion, we assume solar abundances and use main sequence 
temperatures and radii corresponding to
our spectral classifications. These values define absolute Visual
magnitudes, which were compared to the apparent Visual magnitudes and,
after taking into account the measured dilution of the cool companion
(\S \ref{subsec:classification}) and the extinction \citep{sch98},
provided us with the rough distance estimates shown in Table
\ref{tab:parameters}.

We converted positions and motions to Galactocentric
coordinates\footnote{Left--handed Cartesian coordinates where $X$
  increases from Galactic center to anti--Solar direction and $Z$
  increases towards the Galactic North Pole.} ($X$, $Y$, $Z$) and
velocities ($U$, $V$, $W$), adopting a Galactocentric distance of 8.5
kpc for the Sun.  The resulting 6--D phase space coordinates are
presented in Table \ref{tab:6D_coords}.  We also include in this table
the velocity component in the direction of Galactic rotation ($\Theta$),
the component towards the center of the Galaxy ($\Phi$), and the angular
momentum ($I_z$).  The space velocities were transformed to the Local
Standard of Rest by removing a solar motion of (Ð10.0, 5.3, 7.2) km
s$^{-1}$ \citep{deh98}.  We used the {\sc orbit6} code developed by
\citet{ode92} to calculate the orbits of the binaries in the
axisymmetric Galactic potential of \citet{all91}.  In this model, the
disk rotation speed and volume density are 220 km s$^{-1}$ and 0.15
M$_{\sun}$ pc$^{-3}$, respectively.  We reconstructed the path of each
system over a 10 Gyr period with time steps of 1 Myr in order to obtain
a clear picture of the shape of each orbit.  Figure \ref{fig:orbits}
shows the resulting trajectories in the $Y$--$X$, $Z$--$X$, and
$Z$--$\rho$ planes.  The last four columns of Table \ref{tab:6D_coords}
list the apo-- and perigalactic distances ($R_a$ and $R_p$
respectively), the maximum $Z$-direction excursion from the disk, and
the eccentricity ($e \equiv (R_a - R_p)/(R_a + R_p)$), all of which were
determined by integrating the orbits. 

Figure \ref{fig:orbits} reveals the
orbits of PG 1449+653 and PG 1701+359 to be fundamentally different from
those of the other systems.  Whereas PG 1104+243, PG 1317+123, and PG 1338+611
exhibit relatively well--behaved, disk--like trajectories that
never stray far from the Galactic plane (PG 1317+123 being the most
Sun--like), the orbits of PG 1449+653 and PG 1701+359 are significantly
more eccentric and carry these systems much farther out of the disk.
The orbit of PG 1701+359 shows surprisingly little angular momentum,
with a highly--eccentric orbit ($e \sim$ 0.8) bringing it as close as
1.5 kpc to the Galactic center and as far away as 10.6 kpc.  PG 1449+653
currently appears to be near apogalacticon.  We remind the reader that
all of the orbits calculated here represent \textit{approximate}
trajectories; small changes in the initial 6D phase--space coordinates
can significantly change the results, in addition to any encounters with
localized regions of higher density in the Galaxy which are not
accounted for by {\sc orbit6}.

\begin{figure}
  \begin{center}
   \includegraphics[scale=0.85]{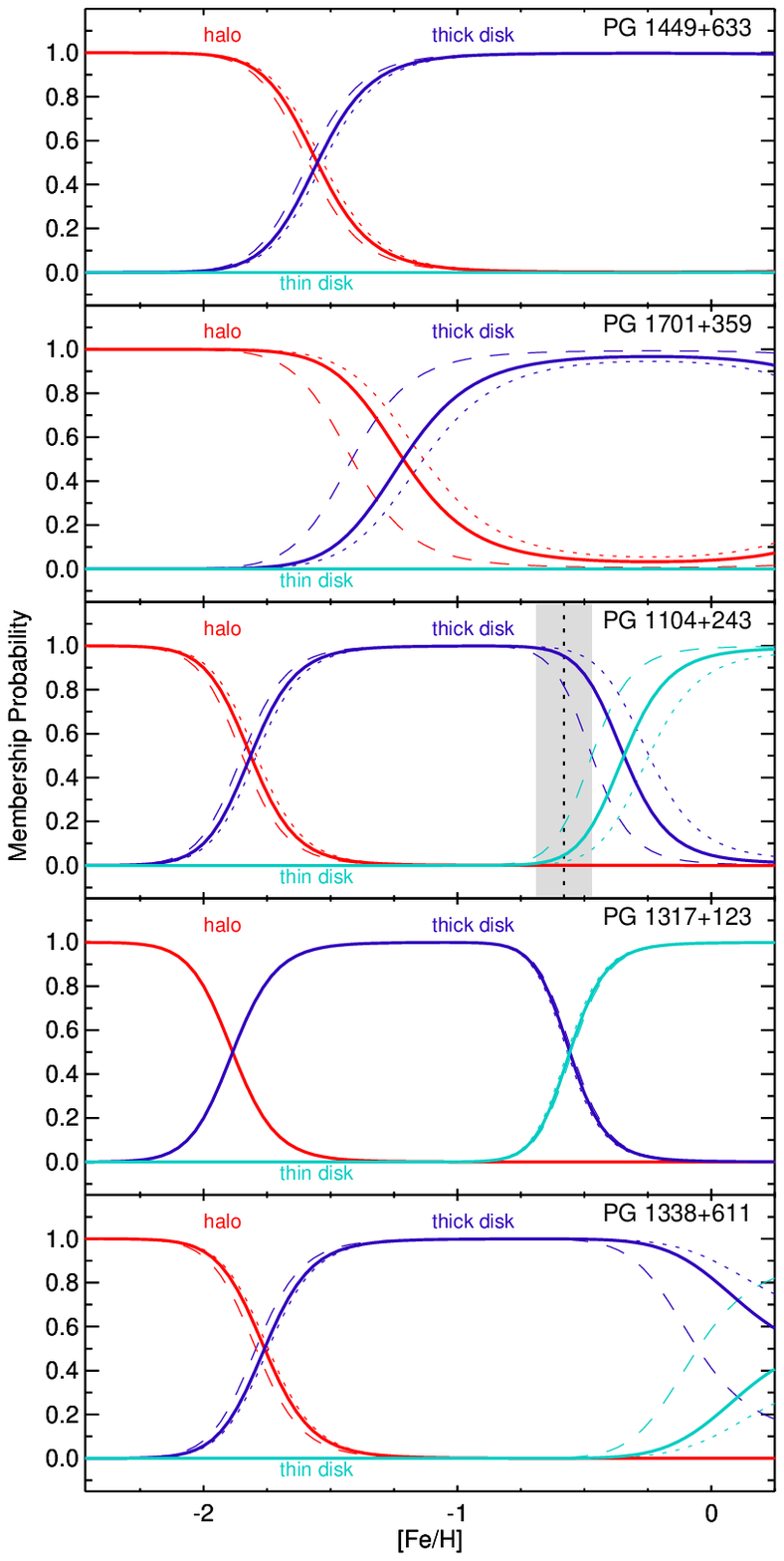}
   \caption{Galactic membership probabilities for PG 1449+653 and PG 
   1701+359 (top two panels).  The solid red, dark blue, and light blue lines denote the 
   probabilities that each system is a member of the halo, thick disk, and 
   thin disk, respectively, for a range of metallicities, using our best estimate 
   for the distance.  The dotted and dashed lines show how these probabilities 
   change when we modify the distance to accommodate the uncertainties shown
   in Table \ref{tab:parameters}. The kinematics alone show it is unlikely either 
   system is a member of the Galactic thin disk.  In the bottom three panels,
   we present the membership probabilities for the three binaries solved in Paper 1.
   The vertical line and shaded region in the PG 1104+243 panel represent the 
   [Fe/H] measurement and associated uncertainty reported by  \citet{vos12}.
    (A color version of this figure is 
   available in the online journal.)}
     \label{fig:pop}
  \end{center}
\end{figure}

We employ the method described in the appendix of \citet{gre07} to
calculate the probabilities that each system belongs to the halo
($P_{\rm halo}$), thick dick ($P_{\rm thick}$), or thin disk ($P_{\rm
  thin}$).  This procedure assumes that all stellar systems belong to
one of these three subpopulations, and that the Galactic velocities
($U$,$V$,$W$) and metallicities ([Fe/H]) of each subpopulation follow
Gaussian distributions defined by the data in Table 4 of \citet{rob03}.
Since we lack precise metallicity estimates for the two systems studied
here, we plot these probabilities in Figure \ref{fig:pop} over a broad
range of metallicities.  

The kinematics alone are enough to rule out thin disk membership for
both PG 1449+653 and PG 1701+359.  Choosing
between the thick disk and halo populations will ultimately require
precise measurements of the metallicity.  If [Fe/H] turns out to be
much less than -1.55 (-1.25), PG 1449+653 (PG 1701+359)
would most likely be a member of the Galactic halo and
not the thick disk.  \citet{vos12} report [Fe/H] = -0.58 $\pm$ 0.11 dex
for PG 1104+243; using this value and the system's kinematics, 
we find that there is $\sim$94\% probability it is
a member of the thick disk.  We cannot assign memberships
 to PG 1317+123 and PG 1338+611 at this time using the data currently available.

The only major 3D kinematical study of hot subdwarfs in the Galaxy was
carried out by \citet{alt04} on a sample of 114 stars selected primarily
from the Hamburg/ESO (HE) and Palomar Green (PG; \citealt{gre86})
surveys; some of their targets were composite--spectra systems.  They
found that the vast majority of sdBs are members of the disk, while only
a minority ($\sim$ 15\%) appear to be halo objects.  Of the disk
population, the thick disk seems to be preferred over the thin disk.  We
cannot make any definitive claims concerning Galactic membership
differences between the composite--spectra systems and the apparently
single or short--period sdB binaries at this time, owing to the small
number of long--period systems currently studied in sufficient detail.
Eventually, a kinematical analysis of a larger sample of
composite--spectra binaries, aided by precise proper motion measurements
from \textit{Gaia}, might provide additional insight into the formation
scenarios of long--period binaries.  As some of these binaries show 
peak--to--peak RV amplitudes in excess of 15 km s$^{-1}$, it is imperative
that such studies use the true systemic velocity when computing
Galactic trajectories, as determined from a well--sampled RV curve
with full phase coverage.


\section{Concluding Remarks}
We have presented follow--up spectroscopic observations of the sdB+G/K
binaries PG 1449+653 and PG 1701+359, the tenth and eleventh such
systems for which orbital parameters have been determined.  Combining
spectroscopic data from the MMT and HET, we find orbital periods around
two years for both binaries.  While we cannot claim an eccentric orbit in the case
of PG 1701+359, PG 1449+653 appears to have a mildly eccentric orbit, 
thereby joining PG 1338+611 (Paper I) and possibly PG 1018+243 \citep{dec12} 
in an emerging group of long--period sdB+/F/G/K binaries with non--circular orbits.
If wide sdB binaries continue to show eccentric orbits, this would suggest one of the following
possibilities:  (1) these systems are or were hierarchical triple--star systems, (2) circularization
was never achieved because the sdB progenitor did not fill its Roche lobe, or (3) 
one of several possible mechanisms pumped eccentricity into the system.
Further details concerning eccentric orbits of sdB+F/G/K binaries and the implications 
thereof are discussed in Paper I.
A kinematical analysis inspired by the
relatively large systemic velocities found for each system shows that
they are probably not members of the Galactic thin disk; we cannot yet
distinguish between thick--disk or halo memberships.

\newpage
Our results continue the trend of finding long periods (on the order of
years) for hot subdwarf stars with G/K-type main sequence
companions.  Binary population synthesis codes (e.g.,
\citealt{han02,han03}) are relatively successful at matching the
observed orbital periods of sdBs with M dwarf or white dwarf companions
but struggle to reproduce the long periods found for sdB+F/G/KV
systems.  Even though relatively few of these long--period binaries have
been studied to date, current observations strongly suggest that (1) the
assumptions made in some hot subdwarf formation scenarios should be
revised and (2) some of the parameterizations in BPS models be
re--tuned.  As demonstrated by \citet{cla12}, subtle variations in the assumptions 
about the minimum core mass required for helium ignition, the efficiency of common envelope ejection,
 the envelope binding energy, the criteria for stable mass transfer, and 
 the amount of mass lost during stable mass transfer can lead to BPS models
 producing significantly different distributions of orbital properties for hot subdwarf binaries.
Orbital parameters for many more sdB+F/G/KV systems must be
measured in order to thoroughly evaluate the predictive success of
current BPS models and determine which parameterizations and assumptions
must be adjusted to match the empirical period distribution.  

\begin{acknowledgements}
  This material is based upon work supported by the National Science
  Foundation under Grant No. AST--0908642.  We thank Rohit Deshpande 
  and Suvrath Mahadevan for providing HRS/HET spectra of MS
  stars and Eva Ziegerer for providing a copy of the ORBIT6 code.   
  The Hobby--Eberly Telescope is a joint
  project of the University of Texas at Austin, the Pennsylvania State
  University, Stanford University, Ludwig--Maximilians--Universit\"{a}t
  M\"{u}nchen, and Georg--August--Universit\"{a}t G\"{o}ttingen. The HET
  is named in honor of its principal benefactors, William P. Hobby and
  Robert E.  Eberly.  This research has made use of NASA's Astrophysics Data 
  System Bibliographic Services and the SIMBAD database, operated at CDS, Strasbourg, France. 

\end{acknowledgements}

{\it Facilities:} \facility{HET (MRS, HRS), MMT (Blue Spectrograph)}


\vspace{4mm}
\begin{thebibliography}{30}
\expandafter\ifx\csname natexlab\endcsname\relax\def\natexlab#1{#1}\fi

\bibitem[{{Allen} \& {Santillan}(1991)}]{all91}
{Allen}, C., \& {Santillan}, A. 1991, RevMexAA, 22, 255

\bibitem[{{Altmann} {et~al.}(2004){Altmann}, {Edelmann}, \& {de Boer}}]{alt04}
{Altmann}, M., {Edelmann}, H., \& {de Boer}, K.~S. 2004, \aap, 414, 181

\bibitem[{{Barlow} {et~al.}(2012){Barlow}, {Wade}, {Liss}, {{\O}stensen}, \&
  {Van Winckel}}]{bar12}
{Barlow}, B.~N., {Wade}, R.~A., {Liss}, S.~E., {{\O}stensen}, R.~H., \& {Van
  Winckel}, H. 2012, \apj, 758, 58 (Paper I)

\bibitem[{{Brown} {et~al.}(2000){Brown}, {Bowers}, {Kimble}, {Sweigart}, \&
  {Ferguson}}]{bro00}
{Brown}, T.~M., {Bowers}, C.~W., {Kimble}, R.~A., {Sweigart}, A.~V., \&
  {Ferguson}, H.~C. 2000, \apj, 532, 308

\bibitem[{{Clausen} \& {Wade}(2011)}]{cla11}
{Clausen}, D., \& {Wade}, R.~A. 2011, \apjl, 733, L42

\bibitem[{{Clausen} {et~al.}(2012){Clausen}, {Wade}, {Kopparapu}, \&
  {O'Shaughnessy}}]{cla12}
{Clausen}, D., {Wade}, R.~A., {Kopparapu}, R.~K., \& {O'Shaughnessy}, R. 2012,
  \apj, 746, 186

\bibitem[{{Cumming} {et~al.}(1999){Cumming}, {Marcy}, \& {Butler}}]{cum99}
{Cumming}, A., {Marcy}, G.~W., \& {Butler}, R.~P. 1999, \apj, 526, 890

\bibitem[{{Deca} {et~al.}(2012){Deca}, {Marsh}, {{\O}stensen}}]{dec12}
{Deca}, J., {Marsh}, T.~R., {{\O}stensen}, R.~H., et~al. 2012, \mnras, 421, 2798

\bibitem[{{Dehnen} \& {Binney}(1998)}]{deh98}
{Dehnen}, W., \& {Binney}, J.~J. 1998, \mnras, 298, 387

\bibitem[{{Geier} {et~al.}(2011){Geier}, {Hirsch}, {Tillich}}]{muchfuss}
{Geier}, S., {Hirsch}, H., {Tillich}, A., et~al.. 2011, \aap, 530, A28

\bibitem[Green et al.(2006)]{gre06} Green, E.~M., Fontaine, 
G., Hyde, E.~A., Charpinet, S., 
\& Chayer, P.\ 2006, Baltic Astronomy, 15, 167 


\bibitem[{{Green} {et~al.}(1986){Green}, {Schmidt}, \& {Liebert}}]{gre86}
{Green}, R.~F., {Schmidt}, M., \& {Liebert}, J. 1986, \apjs, 61, 305

\bibitem[{{Greenstein} \& {Sargent}(1974)}]{gre74}
{Greenstein}, J.~L., \& {Sargent}, A.~I. 1974, \apjs, 28, 157

\bibitem[{{Grether} \& {Lineweaver}(2007)}]{gre07}
{Grether}, D., \& {Lineweaver}, C.~H. 2007, \apj, 669, 1220

\bibitem[{{Han} {et~al.}(2003){Han}, {Podsiadlowski}, {Maxted}, \&
  {Marsh}}]{han03}
{Han}, Z., {Podsiadlowski}, P., {Maxted}, P.~F.~L., \& {Marsh}, T.~R. 2003,
  \mnras, 341, 669

\bibitem[{{Han} {et~al.}(2002){Han}, {Podsiadlowski}, {Maxted}, {Marsh}, \&
  {Ivanova}}]{han02}
{Han}, Z., {Podsiadlowski}, P., {Maxted}, P.~F.~L., {Marsh}, T.~R., \&
  {Ivanova}, N. 2002, \mnras, 336, 449

\bibitem[{{Heber}(1986)}]{heb86}
{Heber}, U. 1986, \aap, 155, 33

\bibitem[{{Johnson}(1966)}]{joh66}
{Johnson}, H.~L. 1966, \araa, 4, 193

\bibitem[Liebert et al.(1994)]{lie94} Liebert, J., Saffer, 
R.~A., \& Green, E.~M.\ 1994, AJ, 107, 1408 

\bibitem[Lucy(2012)]{luc12} Lucy, L.~B.\ 2012, A\&A, 551, 47

\bibitem[{{Lucy} \& {Sweeney}(1971)}]{luc71}
{Lucy}, L.~B., \& {Sweeney}, M.~A. 1971, \aj, 76, 544

\bibitem[{{Mengel} {et~al.}(1976){Mengel}, {Norris}, \& {Gross}}]{men76}
{Mengel}, J.~G., {Norris}, J., \& {Gross}, P.~G. 1976, \apj, 204, 488

\bibitem[Moni Bidin 
\& Piotto(2010)]{mon10} Moni Bidin, C., \& Piotto, G.\ 2010, Ap\&SS, 329, 19 

\bibitem[{{Napiwotzki}(2008)}]{nap08}
{Napiwotzki}, R. 2008, in Astronomical Society of the Pacific Conference
  Series, Vol. 391, Hydrogen-Deficient Stars, ed. A.~{Werner} \& T.~{Rauch},
  257

\bibitem[{{Nelemans}(2010)}]{nel10}
{Nelemans}, G. 2010, \apss, 329, 25

\bibitem[{{Odenkirchen} \& {Brosche}(1992)}]{ode92}
{Odenkirchen}, M., \& {Brosche}, P. 1992, Astronomische Nachrichten, 313, 69

\bibitem[{{{\O}stensen} \& {Van Winckel}(2012)}]{ost12}
{{\O}stensen}, R.~H., \& {Van Winckel}, H. 2012, in ASP Conf. Ser. 452, Fifth Meeting on Hot Subdwarf Stars and
  Related Objects, ed. D.~{Kilkenny}, C.~S. {Jeffery}, \& C.~{Koen}, (San Francisco, CA:  ASP), 163

\bibitem[{{Ramsey} {et~al.}(1998){Ramsey}, {Adams}, {Barnes}}]{ram98}
{Ramsey}, L.~W., {Adams}, M.~T., {Barnes}, T.~G., et~al. 1998, 
Proc. SPIE, 3352, 34

\bibitem[{{Robin} {et~al.}(2003){Robin}, {Reyl{\'e}}, {Derri{\`e}re}, \&
  {Picaud}}]{rob03}
{Robin}, A.~C., {Reyl{\'e}}, C., {Derri{\`e}re}, S., \& {Picaud}, S. 2003,
  \aap, 409, 523

\bibitem[Saffer et al.(2001)]{saf01} Saffer, R.~A., Green, 
E.~M., 
\& Bowers, T.\ 2001, in ASP Conf. Ser. 226, 12th European Workshop on White Dwarfs, ed. J. L. Provencal, H. L. Shipman,
J. MacDonald, \& S. Goodchild (San Francisco, CA: ASP), 408


\bibitem[{{Saffer} {et~al.}(1994){Saffer}, {Bergeron}, {Koester}, \&
  {Liebert}}]{saf94}
{Saffer}, R.~A., {Bergeron}, P., {Koester}, D., \& {Liebert}, J. 1994, \apj,
  432, 351

\bibitem[{{Schlegel} {et~al.}(1998){Schlegel}, {Finkbeiner}, \&
  {Davis}}]{sch98}
{Schlegel}, D.~J., {Finkbeiner}, D.~P., \& {Davis}, M. 1998, \apj, 500, 525

\bibitem[{{Stark} \& {Wade}(2003)}]{sta03}
{Stark}, M.~A., \& {Wade}, R.~A. 2003, \aj, 126, 1455

\bibitem[Tonry \& Davis(1979)]{ton79} Tonry, J., \& Davis, M.\ 1979, \aj, 84, 1511

\bibitem[Vos et al.(2012)]{vos12} Vos, J., {\O}stensen, R.~H., Degroote, P., et al.\ 2012, A\&A, 548, A6 

\bibitem[Wang et al.(2012)]{wan12} Wang, X., Sharon, Wright, 
J.~T., Cochran, W., et al.\ 2012, \apj, 761, 46 

\bibitem[{{Wright} \& {Howard}(2009)}]{wri09}
{Wright}, J.~T., \& {Howard}, A.~W. 2009, \apjs, 182, 205

\bibitem[{{Zacharias} {et~al.}(2012){Zacharias}, {Finch}, {Girard}, {Henden},}]{zac12}
{Zacharias}, N., {Finch}, C.~T., {Girard}, T.~M., {Henden}, A., et~al. 2012, VizieR Online Data
  Catalog, I/322 
\end{thebibliography}

\end{document}